# Geometrization 3.0: The Black Hole Shadow

**Juliano C. S. Neves**

# Geometrization 3.0: The Black Hole Shadow

*Juliano C. S. Neves*
Instituto de Ciência e Tecnologia,
Universidade Federal de Alfenas (Brazil)

**Résumé :** Il y a eu trois géométrisations dans l'histoire. La première est historiquement due à l'école pythagoricienne et à Platon, la deuxième provient de Galilée, Kepler, Descartes et Newton, et la troisième géométrisation de la nature commence avec la relativité générale d'Einstein. Ici, le terme géométrisation de la nature signifie la conception selon laquelle la nature (dans ses différentes acceptions) est largement décrite à l'aide de la géométrie. Dans cet article, je me concentre sur la troisième géométrisation, dans laquelle le phénomène de l'ombre du trou noir relie la forme à la dynamique. Par conséquent, la symétrie de l'espace-temps pourrait jouer le rôle de cause formelle dans la physique des trous noirs. La symétrie de l'espace-temps en tant que cause formelle de l'espace-temps pourrait constituer un point intéressant dans le débat cinématique-dynamique de la théorie de la relativité.

**Abstract:** There have been three geometrizations in history. The first is historically due to the Pythagorean school and Plato while the second comes from Galileo, Kepler, Descartes and Newton, and the third geometrization of nature begins with Einstein's general relativity. Here the term geometrization of nature means the conception according to which nature (with its different meanings) is broadly described by using geometry. In this article, I focus on the third geometrization, in which the black hole shadow phenomenon relates shape to dynamics. As a consequence, spacetime symmetry could play the role of the formal cause in black hole physics. Spacetime symmetry as a formal cause of spacetime could be an interesting point in the kinematics-dynamics debate in the theory of relativity.

# 1 Introduction

According to Kahn, scientists will even today be considered Pythagoreans “if they believe that the laws of nature must be mathematical in form, and that the simpler and more general the mathematical relation is found to be, the more deeply it will penetrate into the nature of things” [Kahn 2001, 172]. The Pythagorean philosophy somehow started a very important movement in the West that is developed in Platonic philosophy. The Pythagorean movement, according to Kahn [2001, 4], is defined as the belief in (i) the immortal soul that concerns a higher form of life and (ii) the mathematical understanding of the world. Therefore, in that regard, today scientists are Pythagoreans if they follow the second point that defines the movement. Thus, according to Kahn [2001, 172], a scientist is Pythagorean only in a “metaphorical sense.” In this article, I focus on the mathematical understanding of the world, in particular on the geometrical worldview which is common to ancient philosophers, pioneers in modern science and today’s physicists. This form of worldview is commonly called the geometrization of nature, in which reality is grounded (explained or described) on geometric concepts. In this way, the geometrization of nature is the act of describing nature using the concepts of geometry. Whether or not nature is ontologically built from geometric concepts, nature or the phenomena (in Kantian sense) is thought about and known mainly from the perspective of geometry in the geometric attitude. If, according to Kant [1998, B 75], the understanding is “the faculty for thinking of objects,” I argue that the *geometric understanding* is the most prominent in the geometric attitude.

It is worth pointing out that the notion of nature for ancient philosophers like Plato or Aristotle is radically different from its modern counterpart that stems from Descartes. For example, for Aristotle nature is process and cause of motion (*Physics* 192*b*8 − 193*b*18 and *Metaphysics* 1014*b*15). As we will see, for Descartes and even Newton, nature conceived of as the material world is a substance without any active principle.

It is accepted that the history of geometrized nature began in the ancient Greek period. Plato was a Pythagorean insofar as the founder of the Academy also believed in the two principles of the Pythagorean school mentioned above. As I said, the point I will focus on in this article is the second—the mathematical understanding or the geometric worldview. In this regard, Plato [1931, 53*b*] accepted the four elements (earth, water, air and fire) represented by the Platonic solids from which the world is made up, but with triangles underneath the elements, i.e., each element would come out of a combination of triangles. This is the first type of geometrization, which I propose to call *geometrization 1.0*.

Plato's cosmology and physics as presented in *Timaeus* is a mathematical worldview,[1] insofar as it is a world built from triangles, or, more specifically, a two-triangle cosmology based on two right triangles [Brisson & Ofman 2020]. Plato's description of the origin of the cosmos stresses the importance of the god Demiurge, or its intellect [*nous*], which is a kind of principle of order in the Platonic cosmology. Starting from a "chaotic matter" governed by necessity [*ananke*], the Demiurge gave a triangular structure to that matter. In this way, triangles are the bricks of the world in the Platonic interpretation, and the material world carries an organizational principle (*nous* or intellect), which in turn means a teleological interpretation of nature. This very point, the teleological issue, is the main difference between ancient and modern physics, which begins with Descartes and others.

The second geometrization or *geometrization 2.0* arose from Galileo's works,[2] the Kepler's laws and the mechanism of Descartes and Newton. Regarding the mechanism, Descartes introduced a new worldview in his book *The World*, where a pure mechanical interpretation of the phenomena is presented. The mechanistic physics of Descartes is opposed to the teleological physics of Plato and Aristotle, as the final cause was banned in modernity. Even so, following the ancient philosophy, Descartes also believed in elements, but in his case just three—earth, air and fire. It is worth emphasizing that, like the ancient philosophers, the name of the elements is just due to a metaphorical usage. For example, Descartes says:[3]

> one can see the difference between this flame, or everyday fire, and the element of fire I have described. And you must also recognise that the elements of air and earth—that is, the second and third element—are not more like the gross air we breathe or the earth on which we walk [...]. [Descartes 2004, 28]

But for the purpose of this article, the most important feature of the material world is what Descartes called extension—a property he identified as the "true form and essence" of matter [Descartes 2004, 36]. The father of analytic geometry conceived of extension alone as the property of bodies. Thus, Descartes reduced matter to extension.[4]

As is well known, Descartes divided the human world into two substances: extension (or matter) and soul. His meditations [Descartes 2017], which also started modern philosophy, are grounded on a geometrical view of the material world. In this regard, Descartes influenced all the thinkers who came after

1. See [Zeyl & Sattler 2019] for an introduction to Plato's cosmology. Also see [Carone 2005] and [Neves 2021*a*,*b*] for aspects of the Platonic cosmology.

2. See [Koyré 1943], and [Machamer & Miller 2021] for an introduction to Galileo's mathematical world.

3. Like commentators on Descartes, I have adopted the page number of the standard edition of Charles Adam and Paul Tannery in quotations.

4. See [Slowik 2021] for an introduction to Descartes' physics and [Jammer 1954] on the Cartesian geometrical worldview.

him, and even Newton [1846] simplified the material world (*à la* Descartes) when he attributed just one property to it—mass. Therefore, modern science is not different from ancient science just because of a "scientific method". It is different mainly because of the conception of the material world. Modern science removed from the material world any hidden principle (like the Platonic principles *nous* and *ananke*), any vestige of soul, intention, teleology or, as Aristotelian scholars called it, any final cause. The teleological cosmos (ordered, for example, by the Demiurge in Platonic cosmology) was abandoned in modernity, and this attitude is one of the two new attitudes characterizing modern science according to Koyré [1943, 403], namely "(1) the destruction of the cosmos [Greek cosmos] and therefore the disappearance in science of all considerations based on that notion; (2) the geometrization of space [...]" or the replacement of the Greek hierarchical space with the homogeneous and isotropic space of the Euclidean geometry. Thus, following Koyré in the same passage:

> these two characteristics may be summed up and expressed as follows: the mathematization (geometrization) of nature and, therefore, the mathematization (geometrization) of science. [Koyré 1943, 403]

Nonetheless, the inert material world or lifeless bodies of modernity "demanded" new properties or principles, which mainly originated during the 20th century because of both quantum mechanics and general relativity. Contrary to the ancient worldview or the Aristotelian causes (also called principles), the modern science of Descartes and Newton considers just efficient and material causes. The former is given by forces, while the latter is inert matter properly speaking. The other two causes (formal and final causes) are ruled out.[5] As I said, the final cause and the teleological interpretation are avoided in modernity. This opinion is shared by great philosophers, and even Nietzsche [2002, § 13]—one of the most important critics of modernity—agrees with modern thinkers as he writes: "watch out for superfluous teleological principles!" But 20th century physics posits that the material world is more sophisticated than just extension or mass, and new properties were necessary in order to better describe or know the world. The most important new property of the objects is particle spin, which comes from quantum mechanics. The material world and the objects (compared to Descartes' or Newton's worldview) present more properties today. But even so, phenomena like quantum entanglement seem to "demand" new properties (arguably, not due to Einsteinian hidden variables).

But what is a cause for Aristotle? In *Physics* 194*b*25, Aristotle says "the form or the archetype, i.e., the definition of the essence, and its genera, are

5. For studies on the final cause in Aristotle's philosophy, see [Gotthelf 1976] and [Angioni 2011], who considers the final to be the most important. For a review on Aristotle's theory of causes, see [Hocutt 1974], [Angioni 2011], and [Falcon 2022].

called causes." Following Hocutt in an interesting article, cause for Aristotle means because:[6]

> Aristotle explicitly presents causes as answers to the question 'Why?', a question which is normally answered by saying 'Because'. [Hocutt 1974, 387].

In particular, in *Metaphysics* 1013*a*25, Aristotle writes that a cause (that which we call formal cause) could be "the form or pattern, i.e., the formula of the essence" of a thing. If the final cause is something dated today, the formal cause is not. As we will see, black hole physics brings the formal cause back to the discussion again. Spacetime symmetry is a kind of formal cause *because* it gives the "essence" or pattern shared by similar spacetimes. For example, a class of spherically symmetric spacetimes shares special vector fields (Killing vectors, as we will see) that define spherical symmetry and conserved quantities. Therefore, conserved charges, like energy and angular momentum, are derived for *all members* of that class.

In particular, symmetry and dynamics are joined in black-hole physics, but this is not something new, because the Noether theorem [Noether 1918] relates symmetry and dynamics from the Lagrangian or from the field theory description (not from the geometric description, which is the point studied here). This means that a given symmetry of the studied system in the Lagrangian formulation of a physical phenomenon, according to the Noether theorem, is translated into a conserved charge and vice versa. However, the point here is the geometrical description and, consequently, symmetry conceived of as a formal cause that could be translated into a dynamical variable. Also, the emphasized thesis is the *degree* of geometrization promoted by the general theory of relativity, in which, for example, the shape of the black hole shadow tells us about the existence of rotation, regardless of the matter content of the black hole. As we will see, such a thesis could play an important role in the kinematics-dynamics debate in relativity.

The third movement towards the geometrization of nature begins with Einstein [1916] and general relativity, a movement I call *geometrization 3.0*. However, following Lehmkuhl [2014], and contrary to the common opinion in physics, it is worth emphasizing that Einstein did not consider general relativity to be the theory that geometrizes gravity. But, as we will see, general relativity brings a lot of geometrical concepts to describe the phenomena. As I mentioned, the point here is the geometric attitude, rather than the ontological discussion about the geometric nature of the world.

Nevertheless, among researchers, the theory of general relativity is commonly referred to as a geometric theory of gravitation. For example, the concept of force, on which Newton's gravity depends, is almost excluded in Einstein's theory of gravity. The geodesic motion in curved spacetimes

6. The debate on the meaning of *aitia* as cause or because is found in several works. See [Vlastos 1969], [Hocutt 1974], [Moravcsik 1974], and [Stein 2011].

describes gravity in general relativity regardless of the concept of force.[7] According to Jammer [1999, 260], for gravity, "Descartes' program has finally been carried out by Einstein." That is, the description of the gravitational phenomenon without action at a distance was achieved in the general relativity context by using the concept of spacetime curvature.

The theory of general relativity was tested during the last century and precise measurements in latter years have confirmed several predictions of Einstein's theory. The detection and observation of both gravitational waves [Abbott, Abbott *et al.* 2016] and the black hole shadows [Akiyama, Alberdi *et al.* 2019*a*, 2022] of Messier 87* (M87*) and Sagittarius A* (Sgr A*) can be described from the general relativity equations. In this article, the main point concerning general relativity is not only the geometrical description of the gravitational phenomenon, but the huge capability of this worldview to translate geometric variables into dynamical ones and vice versa. As we will see, the black hole shadow—phenomenon recently observed for Sgr A*, our central black hole in the Milky Way galaxy, and for M87*, the central black hole in the Messier 87 galaxy—regards a spacetime region whose shape depends on the black hole rotation. Shape and a dynamical variable, namely angular momentum, are related together in this level of geometrization even without any consideration for the material constitution of the body, which is a black hole in the shadow phenomenon. Indeed, geometry and dynamics are already related in the Kepler [1992] second law, where areas are linked to intervals of time of the orbital motion. However, in black hole physics, the geometrization of nature is extended and the *geometric understanding* is applied to a larger number of phenomena.

This article is structured as follows: in Section 2 general concepts of the theory of general relativity are presented in order to illustrate the geometrical features of the theory. In Section 3, the black hole shadow is pointed out as a phenomenon in which the geometric attitude is taken to the highest level. Section 4 provides a brief comment on the relationship between spacetime symmetry, shadow, and the concept of formal cause. In Section 5, the kinematics-dynamics debate is briefly reviewed, and the interpretation of symmetries as formal cause of spacetime is suggested in the debate. Final comments are in Section 6. As this is an article on geometrized nature, here I have adopted geometrized units (or Planck units), i.e., $G = c = k_B = \hbar = 1$, in which $G$ is the gravitational constant, $c$ is the speed of light in vacuum, $k_B$ is the Boltzmann constant, and $\hbar$ is the Planck constant divided by $2\pi$. Thus, distances and the black hole rotation parameter are given in terms of mass in the geometrized units.[8]

7. Jammer [1999, vi] also defends that "what one calls the 'four fundamental forces of nature' are no longer 'forces' in the traditional sense. In short, just like general relativity, modern particle physics seems to support the thesis that the concept of force has reached the end of its life-cycle [...]".

8. See [Wald 1984, appendix F] for more details on the geometrized units.

# 2 The general relativity case

It is commonly stated that the theory of general relativity is a geometric theory of gravity.[9] By conceiving of the gravitational phenomenon as spacetime curvature, general relativity says that the "natural path" of bodies—when there are no other phenomena than gravity—is a geodesic curve. And if there is matter or if spacetime is not empty, the geodesics will be curved paths. Following the pioneers of modern science, who replaced the Greek notion of hierarchical space with the Euclidean space in order to geometrize physics, Einstein replaced the flat Euclidean space with a pseudo-Riemannian manifold to promote a new theory of gravity.

According to general relativity, given a tangent vector $v^\mu$ to a trajectory in spacetime, the tangent vector of a geodesic ($v^\mu = dx^\mu/d\lambda$) is parallel-transported along that special curve. Then the geodesic equations are defined as

$$\frac{d^2x^\mu(\lambda)}{d\lambda^2} + \Gamma^\mu_{\nu\alpha}\frac{dx^\nu(\lambda)}{d\lambda}\frac{dx^\alpha(\lambda)}{d\lambda} = 0, \tag{1}$$

where $\lambda$ is the affine parameter. In time-like geodesics, which are trajectories of massive bodies, $\lambda$ plays the role of the proper time of an observer through that curve. In spacetime, the trajectory of a specific body (whether a massive or massless particle) along a geodesic, indicated as $x^\mu(\lambda)$, is the solution of Eq. (1) and depends on the spacetime metric or spacetime geometry, $g_{\mu\nu}$, which appears in the definition of the affine connection $\Gamma^\mu_{\nu\alpha}$:

$$\Gamma^\mu_{\nu\alpha} = \frac{1}{2}g^{\mu\beta}\left(\frac{\partial g_{\alpha\beta}}{\partial x^\nu} + \frac{\partial g_{\nu\beta}}{\partial x^\alpha} - \frac{\partial g_{\nu\alpha}}{\partial x^\beta}\right). \tag{2}$$

Thus, we need to know the spacetime geometry or metric to evaluate the motion of bodies.

Indeed, general relativity geometrization is also clear, for example, when we read the gravitational field equations,[10] (the Einstein field equations) which are explicitly written as

$$R_{\mu\nu} - \frac{1}{2}Rg_{\mu\nu} = 8\pi T_{\mu\nu}, \tag{3}$$

where $R$ and $R_{\mu\nu}$ are the Ricci scalar and the Ricci tensor, respectively (both are given by the metric tensor $g_{\mu\nu}$), and $T_{\mu\nu}$ is the energy-momentum tensor, something that codes the matter and energy content. Thus, the left side of the Einstein field equations is about geometry, and the right side regards matter and energy. The Einstein equations gave rise to the famous quotation included

9. As we have seen in Introduction, that was not Einstein's interpretation, but it was adopted by many physicists later. See [Lehmkuhl 2014].

10. As we can see, this version of the Einstein field equations is without the cosmological constant term, namely $\Lambda g_{\mu\nu}$.

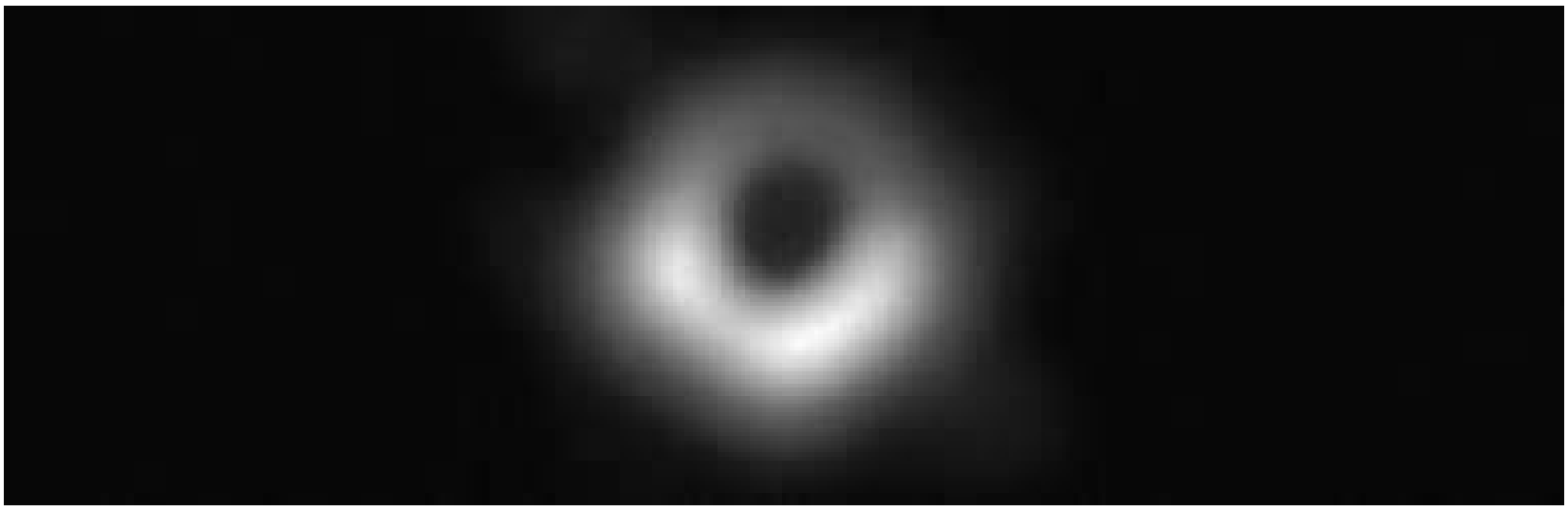

Figure 1: The shadow of M87* at the centre of the Messier 87 galaxy imaged by the EHT collaboration [Akiyama, Alberdi *et al.* 2019*a*]. The black hole shadow is, properly speaking, the interior dark region. Image available on https://eventhorizontelescope.org/press-release-april-10-2019-astronomers-capture-first-image-black-hole. Credit: EHT Collaboration.

in Misner, Thorne & Wheeler's book *Gravitation*: "space tells matter how to move" and "matter tells space how to curve" [Misner, Thorne *et al.* 1973, 5]. In short, for many physicists, this is the simplest reading of general relativity as a geometric theory of gravity and the first step into *geometrization 3.0.*

# 3 The black hole shadow

A deeper level of geometrization in general relativity can be found in black hole physics, especially in the black hole shadow phenomenon. Another can be read from black hole thermodynamics, where, according to the area theorem, the black hole entropy $S$ is proportional to the event horizon area $A$ of a stationary black hole, that is, $S = A/4$ [Wald 2001]. Thermodynamics and geometry are related in that formula, altough black hole thermodynamics is still a speculation in recent theoretical physics. Black hole thermodynamics arises from the black hole dynamics as an analogy, but until now is a set of propositions without strong empirical support. However, the level of geometrization is astonishing, for the geometric concept of area is translated into the thermodynamic concept of entropy.

On the other hand, the black hole shadow presents a greater degree of scientificity when compared to black hole thermodynamics. The image from the centre of the Messier 87 galaxy [Akiyama, Alberdi *et al.* 2019*a*], indicating a supermassive black hole, also called M87*, was arguably the most commented scientific image in 2019. This image (see Fig. 1) shows a dark region or, as the Event Horizon Telescope (EHT) collaboration says [Akiyama, Alberdi *et al.* 2019*a*, 9], a "brightness depression" surrounded by a very bright emission ring. The dark region is actually the shadow of M87*, and its silhouette could be a little deformed when compared to a circle. The bright ring mentioned is called an accretion disk which is exterior to the black hole and the shadow. The accretion disk is composed of matter (plasma) that either orbits or falls into the black hole. The first image of a black hole is an incredible achievement. The EHT collaboration based on a global network of radio telescopes used the very-long-baseline interferometry technique to bring us the image of a supermassive black hole with about 6.5 billion solar masses and 55 million light-years from Earth.

But here let us focus on the shadow phenomenon without considering the influence of the accretion disk on the shadow shape. The black hole shadow is generated by the photon sphere, which is a set of unstable orbits (geodesics) around and exterior to the black hole, that is to say, the photon sphere radius is larger than the event horizon radius. Photons (massless particles) on the photon sphere can either fall into the black hole or escape to "infinity". Part of those photons escapes to the observer position outside the black hole. For a distant observer, when she/he projects the photon sphere onto the Cartesian plane, there will be a dark region that is larger than the black hole, and this larger region is the black hole shadow.

It is worth emphasizing that the procedure indicated below for obtaining the shadow shape is idealized, insofar as, black holes will be considered in vacuum, without an accretion disk like the one indicated in Fig. 1. Such a condition could be acceptable for some configurations of the accretion flow, when the shape and size of the shadow surrounded by an accretion disk will be almost identical to the vacuum case.[11]

## 3.1 Static black holes

Static black holes are non-rotating black holes. This class of solutions of the Einstein field equations is spherically symmetric, or, in other terms, spacetime symmetry is the spherical symmetry. A rigorous definition of spherical symmetry is given in terms of Killing [1892] vector fields, and the relation between those special vector fields and conserved charges will be mentioned in Section 4.

Due to the gravitational lensing, the observed radius of the shadow is larger than the photon sphere radius. For example, for a static spacetime like the Schwarzschild [1916] black hole—which today is referred to as a black hole solution of the Einstein field equations (3) without either rotation or charge (whether electric or magnetic charges)—the spacetime metric of the corresponding solution in the $(t, r, \theta, \phi)$ coordinates is written as

$$ds^2 = -\left(1 - \frac{2M}{r}\right) dt^2 + \frac{dr^2}{\left(1 - \frac{2M}{r}\right)} + r^2 \left(d\theta^2 + \sin^2\theta d\phi^2\right), \tag{4}$$

11. See [Perlick & Tsupko 2022] for some comments on this point. By considering a vacuum spacetime (without an accretion disk or matter between the shadow and the observer) light rays travel on geodesic curves.

where $M$ stands for the black hole mass, and the photon sphere has the following radius: $r_{\rm ph} = 3M$ (measured from the centre of the black hole) or $r_{\rm ph} = 3GM/c^2$ in the International System of Units. On the other hand, according to Perlick & Tsupko [2022], the shadow radius of the Schwarzschild black hole is, after approximations, given by[12]

$$r_{\rm Sch} = \frac{r_{\rm ph}}{\sqrt{1-\frac{2M}{r_{\rm ph}}}}, \tag{5}$$

which is constant. For the static Schwarzschild black hole, the photon sphere is a perfect sphere and thus the shadow silhouette is a perfect circle, when projected onto the Cartesian plane by a distant observer.

## 3.2 Rotating black holes

However, astrophysical black holes like M87* or Sgr A* are not static or non-rotating black holes. Arguably, they have rotation or angular momentum, even though such a parameter has yet to be precisely measured.[13] The most common mathematical description of an astrophysical black hole is called the Kerr metric or spacetime [Kerr 1963], which is also a solution of Einstein's equations (3) and describes the vacuum exterior region of a rotating and uncharged mass. The Kerr metric is not spherically symmetric, its symmetry is the axial symmetry, thus the Kerr spacetime is symmetric around the rotation axis of the black hole. The Kerr metric, in the same $(t, r, \theta, \phi)$ coordinates, reads

$$\begin{aligned} ds^2 = &-\left(1-\frac{2Mr}{\Sigma}\right)dt^2 - \frac{4Mar\sin^2\theta}{\Sigma}dtd\phi + \frac{\Sigma}{\Delta}dr^2 + \Sigma\, d\theta^2 \\ &+ \frac{\sin^2\theta}{\Sigma}\left[(r^2+a^2)^2 - \Delta a^2\sin^2\theta\right]d\phi^2, \end{aligned} \tag{6}$$

where

$$\Delta = r^2 + a^2 - 2Mr \quad \text{and} \quad \Sigma = r^2 + a^2\cos^2\theta. \tag{7}$$

Here just like the Schwarzschild metric, $M$ is related to the black hole mass, and the new parameter in Eq (6), namely $a$, is the rotation parameter, which is given by the black hole angular momentum ($J$) per unit mass: $a = J/M$. With $a = 0$, the Kerr metric tuns into the Schwarzschild spacetime (4), in that there is no longer any rotation. It is worth emphasizing that, in order to get a compact object with an event horizon (like a black hole), the

12. The first calculation of the Schwarzschild shadow was made by [Synge 1966]. And the first representation of the Schwarzschild black hole surrounded by an accretion disk was achieved by Luminet [1979].

13. See [Fragione & Loeb 2020, 2022], [Nemmen 2019], and [Tamburini, Thidé *et al.* 2019] for some estimations for the black hole rotation parameter of either black holes.

rotation parameter must lie in the interval $0 \leq a \leq M$. For $a > M$, there would be the so-called naked singularity or a singularity unprotected by an event horizon, which is the one-way membrane delimiting the interior and the exterior of a black hole. For the shadow phenomenon, even naked singularities, wormholes, gravastars or compact objects without an event horizon cast shadows.[14] The event horizon is not relevant for the shadow calculations, because the event horizon radius $r_{\rm h}$ of the Kerr black hole is smaller than the photon sphere radius $r_{\rm ph}$. That is, as I said, the shadow phenomenon is outside the black hole or $r_{\rm ph} > r_{\rm h}$.[15]

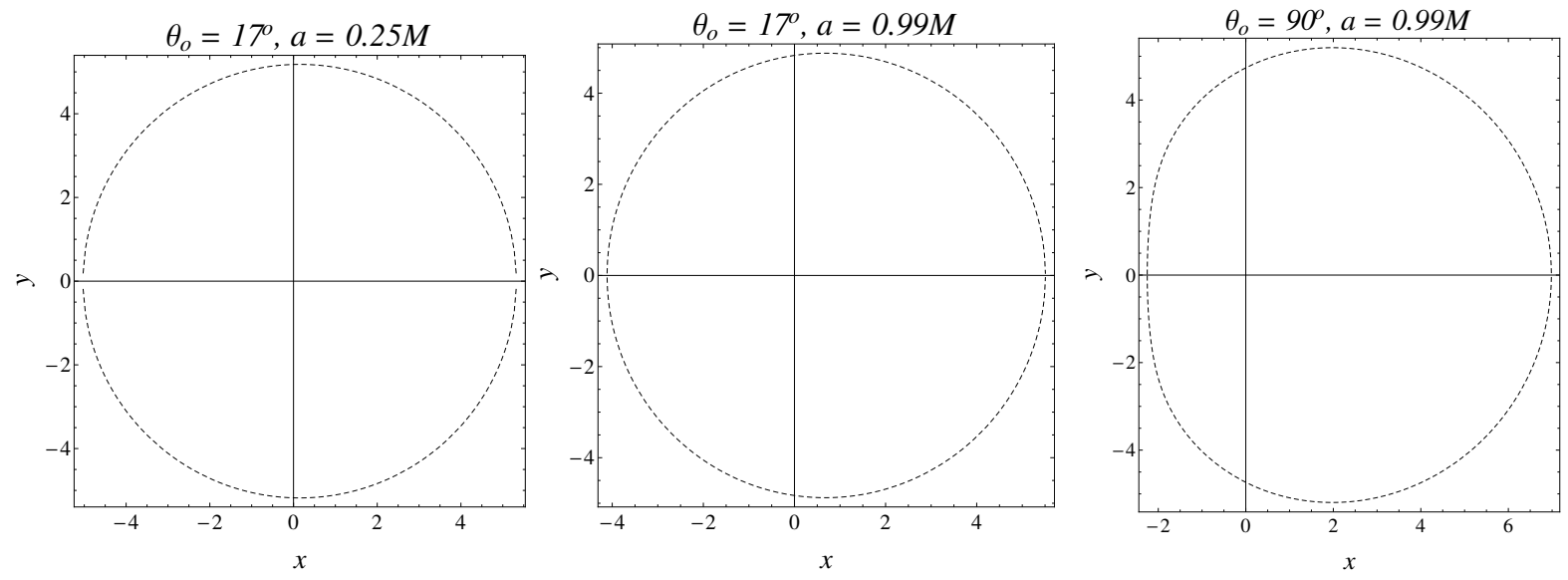

Figure 2: The shadow silhouette for the Kerr black hole for different values of observation angle, $\theta_o$, and rotation parameter, $a$. The value $\theta_o = 17°$ is arguably our observation angle in relation to the rotation axis of M87* [Akiyama, Alberdi *et al.* 2019*a*]. Here, the black hole is rotating counter-clockwise. As we can see, the deviation from circularity of the shadow is more evident for both large values of $\theta_o$ and $a$. The maximum value of $\Delta C$ occurs for $\theta_o = 90°$, i.e., the observer is on the black hole equatorial plane. For $a = 0$, the shadow silhouette is a perfect circle independently of the observation angle. The $x$ and $y$ axes are in geometrized units.

For rotating black holes, like the Kerr black hole, Eq. (5) is no longer valid, because the shadow of rotating black holes could have a deformed shape or a "D" shape, that is the shadow radius is not constant. As we can see in Fig. 2, the deformation of the shadow silhouette also depends on the observation angle, but the *conditio sine qua non* for the deformed shape is the black hole rotation. Thus, the photon sphere radius now depends on the black hole rotation parameter. As the hole rotates, spacetime dragging will occur, and

14. Naked singularities, wormholes and gravastars are hypothetical exotic solutions of the gravitational field equations. As I mentioned, they have no event horizon. Regarding the shadow shape, static black holes and static exotic solutions cast circular shadow. As we will see, rotation alone produces the "D" shape in the shadow phenomenon regardless the matter content.

15. For comparison, as we have seen, for the Schwarzschild black hole $r_{\rm ph} = 3M$, but the event horizon radius is $r_{\rm h} = 2M$. If we consider the accretion disk, its orbit will be given by the ISCO (Innermost Stable Circular Orbit) localization, which in the Schwarzschild black hole is $6M$, outside the photon sphere as we see in Fig. 1.

cause the "D" shape of the shadow. Looking at the shadow silhouette like Fig. 2, photons on the left side of the shadow silhouette are travelling in the same direction as the black hole rotation, but photons on the right side of the silhouette are travelling in the opposite direction to the black hole rotation.

In order to obtain the photon sphere orbits, the radial component of the geodesic equations (1) for photons need to be solved. For the Kerr spacetime (6), the radial component is given by

$$\frac{dr}{d\lambda} = \frac{\sqrt{\mathcal{R}(r)}}{\Sigma}, \tag{8}$$

where $\mathcal{R}(r)$ is some sort of effective potential. The shadow silhouette, or the curves that delimit the photon sphere, comes from the following conditions for the radial component: (i) $\mathcal{R}(r_{\text{ph}}) = 0$ and (ii) $d\mathcal{R}(r_{\text{ph}})/dr = 0$. The details of the function $\mathcal{R}(r)$ are available in [Carter 1968], where the geodesics for the Kerr spacetime were calculated for the first time. The two conditions give us the photon sphere radius $r_{\text{ph}}$ for the Kerr black hole.

Following [Bardeen 1973], where the equations for shadows in the Kerr spacetime were explicitly shown for the first time, the idea is to project the shadow onto the Cartesian plane like Fig. 2 in order to compute the deviation from circularity of the shadow. The Cartesian components $x$ and $y$ of the celestial coordinates of the shadow silhouette (measured by a distant observer) for a Kerr black hole are

$$x = \frac{\xi}{\sin\theta_o}, \tag{9}$$

$$y = \pm\sqrt{\eta + a^2\cos^2\theta_o - \xi^2\cot^2\theta_o}\ , \tag{10}$$

where the angle $\theta_o$ is the observation angle, which is the angle given by the observer position in relation to the black hole rotation axis. When $\theta_o = 90°$, it means that the observer is located on the equatorial plane of the black hole. The parameters $\xi = L/E$ and $\eta = K/E$ are constants for each photon that travels on the photon sphere, that is to say, these parameters are conserved quantities for photons passing through the photon sphere. The conserved charges for each photon are: $L$, the angular momentum, $E$, the energy, and $K$, the Carter constant, which is an integration constant. For the Kerr black hole, the parameters $\xi$ and $\eta$ could be written in terms of the two black hole parameters, namely $M$ and $a$. Explicitly, one has [Cunha & Herdeiro 2018]

$$\xi = -\frac{(r_{\text{ph}} - 3M)\, r_{\text{ph}}^2 + (r_{\text{ph}} + M)\, a^2}{(r_{\text{ph}} - M)\, a}, \tag{11}$$

$$\eta = \frac{\left(3r_{\text{ph}}^2 + a^2 - \xi^2\right) r_{\text{ph}}^2}{r_{\text{ph}}^2 - a^2}. \tag{12}$$

The Cartesian components of the shadow silhouette (9)–(10) depend on the photon sphere radius, because $\xi(r_{\text{ph}})$ and $\eta(r_{\text{ph}})$ according to the above equations. Like the spherical case, $r_{\rm ph}$ is the photon sphere radius and, contrary to non-rotating black holes, $r_{\rm ph}$ is not constant for their rotating counterparts. The photon sphere radius is given within the range $r_{\rm ph-} \leq r_{\rm ph} \leq r_{\rm ph+}$ for rotating black holes.

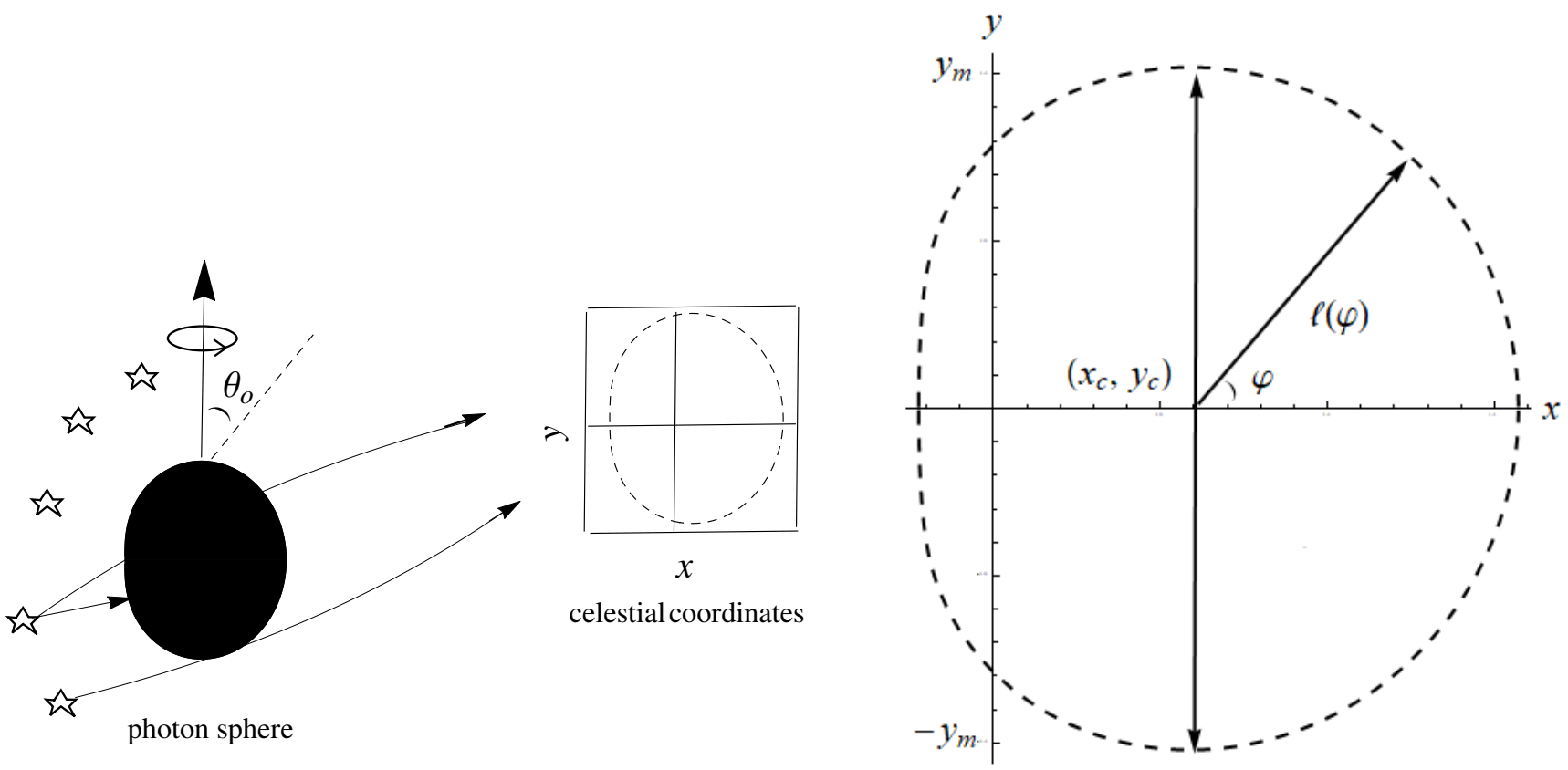

Figure 3: On the left, the shadow phenomenon (caused by the bending of light) described by a distant observer projecting it onto the Cartesian plane with the aid of the celestial coordinates $(x, y)$. On the right, a schematic representation of the black hole shadow, whose silhouette is indicate by the dashed line.The shadow radius $\ell(\varphi)$ is not constant for rotating black holes. The "D" shape of the shadow is due to the black hole rotation that causes the spacetime dragging which is absent with static black holes. The coordinates $(x_c, y_c)$ indicate the centre of the shadow, and $\pm y_m$ are maximum and minimum values of the $y$ coordinate. The shadow is symmetric to the $x$-axis.

In order to obtain an equation that relates the shadow silhouette and the black hole rotation, I will follow Bambi, Freese *et al.* [2019]. According to EHT [Akiyama, Alberdi *et al.* 2019*a*], the M87* shadow is $\lesssim 10\%$ away from a perfect circle, which means the shadow deviation from circularity of that supermassive black hole is $\Delta C \lesssim 10\%$. The parameter $\Delta C$ is then constructed as follows: the first step is obtaining the average radius of the shadow $\bar{R}$, which is defined as

$$\bar{R}^2 \equiv \frac{1}{2\pi} \int_0^{2\pi} \ell(\varphi)^2 d\varphi, \tag{13}$$

where the shadow radius reads

$$\ell(\varphi) \equiv \sqrt{(x - x_c)^2 + (y - y_c)^2}. \tag{14}$$

The coordinates $(x_c, y_c)$ provide the geometric centre of the shadow projected onto the Cartesian plane (see Fig. 3 and [Bambi, Freese *et al.* 2019] for details). According to the EHT report [Akiyama, Alberdi *et al.* 2019*a*], the deviation from circularity is given by the root-mean-square distance from the average shadow radius, namely

$$\Delta C \equiv \frac{1}{\bar{R}}\sqrt{\frac{1}{2\pi}\int_0^{2\pi}\left(\ell(\varphi)-\bar{R}\right)^2 d\varphi. \tag{15}$$

In black hole physics, the deviation from circularity $\Delta C$ of M87* has been used to constrain the theory of general relativity [Bambi, Freese *et al.* 2019], [Khodadi, Lambiase *et al.* 2021] and theories beyond the Einsteinian theory [Banerjee, Chakraborty *et al.* 2020], [Neves 2020*b*,*a*], [Kumar & Ghosh 2021, Kumar, Kumar *et al.* 2020], [Vagnozzi & Visinelli 2019]. Such modified or alternative models and theories provide new parameters that are directly constrained from the upper bound $\Delta C \lesssim 10\%$. Even though alternatives to the Kerr spacetime are possible, the working hypothesis of the EHT collaboration was the Kerr metric by assuming previous observations and simulations. According to Weinstein [2021, 41]:

> the EHT collaboration has made an 'inference to the best explanation' [...] to conclude that the data effectively confirm the hypothesis that the object at the centre of M87 is in fact a supermassive Kerr rotating black hole.[16]

It is important to emphasize that the shadow silhouette is independent of the type of matter and energy inside the black hole. In this way, once the back hole exists, the type of ordinary or exotic matter inside the event horizon does not matter (whether considering a supermassive black hole or not). On the one hand, as we can see in Fig. 2, given a black hole with mass $M$, the deviation from circularity $\Delta C$ depends both on the observation angle $\theta_o$ and the rotation parameter $a$. $\Delta C$ is more pronounced for larger values of $\theta_o$. And large deviations are only possible for large rotation parameters. On the other hand—for a given black hole with mass $M$, whose value could be estimated from different approaches and methods and which is observed at a well-known angle $\theta_o$, while also inferred from alternative approaches other than the black hole shadow image—the deviation from circularity just depends on the black hole rotation. Then the shadow shape and the rotation parameter are related. Even before the EHT image, authors [see Takahashi 2004 and Hioki & Maeda 2009] tried to evaluate the black hole rotation from the shadow shape by using other shadow parameters like the oblateness [Tsupko 2017].

16. It is worth noting that alternatives to the Kerr spacetime were studied by the EHT collaboration to fit the M87* data. These options include black holes with additional fields, black holes from alternative theories of gravity and black hole mimickers like wormholes. See [Akiyama, Alberdi *et al.* 2019*b*, Section 7.4].

# 4 Spacetime symmetry as formal cause

As we have seen in Section 3, non-rotating and rotating black holes might cast different shadow silhouettes. Geometrically speaking, the main difference between these two types of objects is the spacetime symmetry. Both the Schwarzschild and the Kerr black holes are stationary black holes. However, a non-rotating black hole referred to as the Schwarzschild spacetime is also conceived of as static spacetime, which is a special class of geometries. Static black holes, like Schwarzschild's, are spherically symmetric. On the other hand, the Kerr black hole is an axially symmetric solution of Einstein's field equations (3), because it is only symmetric around the rotation axis.

In differential geometry,[17] as an alternative to the [Noether 1918] theorem (which concerns a field theory description), the relation between symmetries and conserved quantities is given by vector fields defined by Killing [1892]. Given a spacetime indicated as $(\mathcal{M}, g_{\mu\nu})$, that is, a manifold $\mathcal{M}$ endowed with the metric tensor $g_{\mu\nu}$, the Killing vector fields on $\mathcal{M}$ are in one-to-one correspondence with the continuous symmetries of that spacetime. Along the direction of Killing fields, spacetime is unchanged and such fields then generate so-called isometries, which involves the metric tensor $g_{\mu\nu}$ remaining unchanged along the field direction.[18]

As a consequence from the standpoint of physics and given the tangent vector $v^\mu$ of a geodesic, as defined in Section 2, a Killing vector $\xi^\mu$ yields the product $v^\mu \xi_\mu =$ constant along that geodesic, giving rise to a conserved quantity like energy, momentum or angular momentum for particles along the special curve. This means that the more Killing vector fields there are, the more a spacetime is symmetric and more conserved quantities exist in that spacetime.[19] A Killing vector field is a solution of the Killing equation

$$\nabla_\mu \xi_\nu + \nabla_\nu \xi_\mu = 0, \tag{16}$$

where $\nabla_\mu$ is the covariant derivative operator.

A spacetime written as the Schwarzschild metric presents more Killing vectors than the Kerr spacetime. There are four linear independent Killing vector fields for the Schwarzschild metric and two for the Kerr spacetime. As I have said, both black holes are described by stationary spacetimes, which means the spacetime metrics (4) and (6) do not depend on the time coordinate $t$. In a coordinate-independent way, without explicitly mentioning the coordinate $t$, a given spacetime is stationary if it possesses a time-like Killing vector field, which generates the energy conservation of bodies or

17. See [Wald 1984, appendix 6] for an introduction to the Killing vector fields in general relativity.

18. Technically, an isometry is a diffeomorphism $\phi$ such that $\phi_* g_{\mu\nu} = g_{\mu\nu}$.

19. It is worth pointing out that the solutions studied here, the Schwarzschild spacetime and the Kerr spacetime, are "ideal solutions" or approximations. Astrophysical objects change with time, so they have fewer symmetries.

| **Black Hole Symmetry** | **Rotation Parameter** | **Deviation from Circularity** |
|---|---|---|
| Spherical Symmetry | $a = 0$ | $\Delta C = 0$ |
| Axial Symmetry | $a \neq 0$ | $\Delta C \geq 0$ |

Table 1: The spacetime symmetry of a given black hole, the rotation parameter $a$ (related to the black hole angular momentum), and the deviation from circularity $\Delta C$ of the black hole shadow. It is worth pointing out that there will be finite and $\Delta C \neq 0$ for rotating black holes alone for nonzero values of the observation angle $\theta_o$. On the other hand, for non-rotating black holes, $\Delta C = 0$ is independent of the observation angle.

particles along the geodesics. Moreover, the Schwarzschild spacetime is also a static spacetime. For this type of spacetime, the time reversal transformation $t \to -t$ keeps the metric unchanged. Contrary to Kerr spacetime, there are no cross terms like $dtdx^a$ (for $a = 1, 2, 3$ or $a = r, \theta, \phi$ in the coordinates adopted here) in static spacetimes and the metric remains invariant under a time reversal transformation. As for the time-like Killing field, the Schwarzschild spacetime also possesses a time-like Killing vector field orthogonal to a family of hypersurfaces, but Kerr's does not. These are the reasons why the former is static and the latter is not. In static spacetimes nothing changes, but Kerr includes black hole rotation and the notion of change.

Lastly, as the Schwarzschild black hole is spherically symmetric, it has the same symmetries as the two-sphere $S^2$. In terms of Killing vector fields, a spherically symmetric spacetime admits three spacelike Killing fields $(\xi_1^\mu, \xi_2^\mu, \xi_3^\mu)$ and the following commutation relations [see Carroll 2004, chap. 5]:

$$\begin{aligned} [\xi_1^\mu, \xi_2^\mu] &= \xi_3^\mu, \\ [\xi_2^\mu, \xi_3^\mu] &= \xi_1^\mu, \\ [\xi_3^\mu, \xi_1^\mu] &= \xi_2^\mu. \end{aligned} \tag{17}$$

In terms of symmetry groups, these three fields are generators of the Lie algebra of SO(3), which is the group of rotations. Moreover, these spacelike Killing vectors are related to the conservation of angular momentum. In the Schwarzschild spacetime, angular momentum is conserved around the three axes. However, as the Kerr spacetime only admits the spacelike Killing vector field related to the black hole rotation axis, only the component of the angular momentum around the rotation axis direction is conserved.

As we have seen, both the deformation of the shadow silhouette owing to the black hole angular momentum and spacetime symmetry are directly related in the shadow phenomenon insofar as spacetime symmetry and dynamics are related in this phenomenon. Spherically symmetric and axially symmetric black holes yield different shadows and, consequently, different black hole angular momenta. It is worth mentioning Petitot who considered 20th century physics from a neo-Kantian or transcendental standpoint and traced this very relation between geometry or symmetry and dynamics as follows:

> we have indeed seen that there is a telos of geometrization in physics: transforming principles of symmetry into dynamic principles. [Petitot 1997, 231, my translation]

And this "telos" or attitude towards the development of physics is present once again in general relativity according to the phenomenon of the black hole shadow.

Furthermore, I argue that spacetime symmetry may be conceived of as the formal cause of black holes or, equivalently, of spacetimes. As we have seen in the Introduction, the formal cause, according to Aristotle [*Metaphysics* 1013*a*25], is "the form or pattern, i.e., the formula of the essence" of a thing. Aristotle [*Posterior Analytics* 72*b*5] says that the definition or the statement of essence of something is a self-evident explanation. According to [Hocutt 1974, 388], the material, efficient, formal, and final causes are explanations giving us the *because* of a fact or event in which a cause is an explanation, and

> an explanation is a syllogism, a demonstration in which an *explanandum*, a statement of inherence of an attribute in a subject, is the conclusion, and in which the premises are the *explanans*. [Hocutt 1974, 391]

In order to explain something formally, the chain of explanations should include the last link in which an explanation is self-evident. Therefore, the formal cause or explanation is the last link in the chain of explanations of something or, in other terms, the description of the "essence" of something. Here, from an epistemological perspective, I am using "essence" as an explanation rather than something related to a substance from an ontological perspective.

In this sense, spacetime symmetry gives us the "essence" or explains why spacetime is what it is, and constrains (by using the metric tensor $g_{\mu\nu}$) every phenomenon like the shadow phenomenon. A given spacetime symmetry, like spherical symmetry, gives the pattern or "essence" of any spherically symmetric spacetime in terms of Killing vectors and constrains any possible metric tensor that represents that spacetime. For example, the spherical symmetry and its corresponding Killing vector fields answer the questions "what is the cause of spherically symmetric spacetimes?" or "why is a spherically symmetric spacetime the kind of thing it is?". I argue that spacetime symmetries are the *explanans* and spacetime is the *explanandum.* Thus the spacetime symmetry in general relativity (and, as we will see, in special relativity) is the "essence" of a given spacetime, it is the formal cause or *the because* a family of spacetime metrics is what it is, giving rise to certain types of phenomena. In particular, as we saw, the shadow shape is constrained by the spacetime symmetry. Therefore, one can call the spacetime symmetry (and its corresponding Killing vector fields) the formal cause of a black hole or, equivalently, the formal cause of its corresponding spacetime.[20]

# 5 Symmetries and dynamics

What is the importance of the approach in Section 4 to the recent debate in philosophy of spacetime? For the interpretation of special relativity as either a kinematic theory or a dynamic theory, symmetries and Killing vector fields could help us in this debate.[21]

Let us briefly review this debate. The standard interpretation of special relativity, following Einstein [1905] and Minkowski [1909], conceives of time dilation and space contraction as kinematical effects of the Minkowski spacetime, which means both the relativistic effects and the Lorentz transformations are explained by the Minkowski spacetime and its structure. On the other hand, Brown & Pooley [2001] adopt a dynamical view of special relativity: arguably, the Lorentz transformations (the Lorentz symmetry or the Lorentz invariance in a broader sense) could lead to the Minkowski spacetime. According to the authors, the dynamics of matter from a Lorentz invariant theory (a quantum theory of gravity, for example) are thought to lead to the Minkowski spacetime. However, as we will see, by using the argument of the last section, the Lorentz symmetry explains the Minkowski spacetime.

First of all, Minkowski spacetime is an example of a maximally symmetric spacetime [see Carroll 2004, chap. 3]. For those four-dimensional spacetimes,[22] curvature is constant, and there are ten associated Killing vector fields. For Minkowski geometry, those vectors mean four generators of translations (with the corresponding energy and linear momentum conservation) and six generators of "rotations". Those ten vectors generate the Poincaré group (also known as inhomogeneous Lorentz group). The last six generators are three generators of spatial rotations (with the corresponding angular momentum

20. This is not the first time that the Aristotelian physics and general relativity have been brought together. According to [Jammer 1954, 20], as for the Aristotelian notion of natural places, "the idea of 'geodesic lines', determined by the geometry of space, and their importance for the description of the paths of material particles or light rays, suggest a certain analogy to the notion of 'natural places' [...]". In [Neves 2018], this analogy is emphasized by using the relativistic notion of conformal infinities. Then general relativity once again approaches Aristotelian physics as symmetry plays the role of the formal cause. Interestingly, these analogies would sound strange to Aristotle, because, as is well known, the founder of the Lyceum refused to be a Pythagorean [see Koyré 1943, for this Aristotle's attitude].

21. In this debate, for the kinematics supporters, see [Norton 2008], [Janssen 2009], and [Hagar & Hemmo 2013]. For the dynamics supporters, see [Brown & Pooley 2001], and [Brown & Pooley 2006]. Also, see [Acuña 2016] for the middle ground in this debate. Interestingly, none of those articles mention Killing vector fields.

22. Also de Sitter and anti-de Sitter spacetimes are maximally symmetric spacetimes.

conservation) and three boosts that preserve the spacetime interval. These last six generators form the homogeneous Lorentz group with the Lorentz transformations.[23] As I argued that spacetime symmetry (represented by Killing vector fields) is the formal cause of spacetime—that is, a given set of symmetries explains a class of spacetimes—and the Lorentz transformations are symmetries, we may conclude that all symmetries of Minkowski spacetime, including the Lorentz symmetry, explain the Minkowski spacetime.

In the debate, Janssen, defending the kinematical interpretation, argues that

> the space–time symmetries are the *explanans* and that the Lorentz invariance of the various laws is the *explanandum*, while Brown argues that it is the other way around. [Janssen 2009, 26]

Here I disagree with Janssen, because the Lorentz symmetry is among spacetime symmetries of the Minkowski spacetime. In this sense, the Lorentz symmetry and others explain the Minkowski spacetime. That is, spacetime symmetries are the *explanans*, and the Minkowski spacetime is the *explanandum*. Interestingly, my argument partially supports the Brown & Pooley [2001] thesis because the Lorentz symmetry explains the Minkowski spacetime. Partially, because the Minkowski spacetime is endowed with more symmetries than the Lorentz's, and no reference to matter is made.

The kinematics–dynamics debate is more subtle in general relativity. The notion of kinematics is just local in general relativity. But it is interesting to note how similar the length contraction in special relativity and the black hole shadow silhouette in general relativity are. Both phenomena are matter-constitution independent and are explained fundamentally by the spacetime symmetries.

# 6 Final comments

Black hole physics provides the most radical Pythagorean-like interpretation of the world. The Pythagorean worldview is the philosophical point of view in which mathematics or geometry is essential for understanding the world. As we have seen, there are three steps into the geometrization of nature: the Pythagorean-Platonic, the Galilean-Keplerian-Cartesian-Newtonian, and the Einsteinian. The last step, which I call *geometrization 3.0*, was the focus of this article. Black hole physics, a very special chapter in general relativity, brings out the black hole shadow phenomenon recently observed by the EHT collaboration [Akiyama, Alberdi *et al.* 2019*a*, 2022], which presents the highest level of geometrization ever achieved. The shape of the black hole shadow, which is a geometric quantity, could be translated into a dynamical variable,

23. See [Hawking & Ellis 1973, 62].

namely the angular momentum (or the rotation parameter) of the black hole. An important point here is that the shadow silhouette does not depend on the type of matter-energy content inside the hole.

Different shadow shapes are related to spacetime symmetries. A spherically symmetric black hole only casts circular shadows. On the other hand, axially symmetric black holes or rotating black holes could cast deformed or not perfectly circular shadows. According to Aristotle, the formal cause is some sort of pattern or essence which explains something essentially. As we have seen, spacetime symmetry gives us the pattern or "essence" of spacetime, constraining the metric and, consequently, the physical description and the dynamics. Indeed, spacetime symmetry explains spacetime. Thus, spacetime symmetry could be conceived of as the formal cause of a black hole or its corresponding spacetime in the interpretation presented in this article. Considering symmetry to be the formal cause of spacetime means that Lorentz symmetry explains (alongside other symmetries) the Minkowski spacetime. This could be a fresh argument in the kinematics-dynamics debate on the philosophy of spacetime.

# Acknowledgments

I am grateful to the ICT-UNIFAL for its hospitality and to the Conselho Nacional de Desenvolvimento Científico e Tecnológico (CNPq), Brazil, (Grant No. 170579/2023-9) for the financial support. I also thank two anonymous referees for the interesting points raised during the review process.